\documentstyle[multicol,prl,aps,epsf]{revtex}
\begin{document}
\tightenlines
\title{Quantum Decoherence in Disordered Mesoscopic Systems}
\author{Dmitrii S. Golubev$^{1,2}$ and Andrei D. Zaikin$^{2,3}$}
\address{$^{1}$ Physics Department, Chalmers University of Technology,
S-41296 G\"oteborg, Sweden\\
$^2$ I.E.Tamm Department of Theoretical Physics, P.N.Lebedev
Physics Institute, Leninskii pr. 53, 117924 Moscow, Russia\\
$^3$ Institut f\"{u}r Theoretische Festk\"orperphysik,
Universit\"at Karlsruhe, 76128 Karlsruhe, Germany}

\maketitle

\begin{abstract}
We point out that the low temperature saturation of the electron
phase decoherence time in a disordered conductor
can be explained within the existing theory of weak
localization provided the effect of quantum (high frequency)
fluctuations is taken into account. Making use of the
fluctuation-dissipation theorem we evaluate the quantum decoherence
time, the crossover temperature below which thermal effects become 
unimportant, and the weak localization correction $\delta \sigma$ 
at $T=0$. For 1d systems the latter is found to be $\delta \sigma 
\propto 1/ \sqrt{N}$, where $N$ is the number of conducting channels. 
\end{abstract}


\begin{multicols}{2}

Quantum interference between electrons has a strong impact
on electron transport in a disordered metal, leading
to the so-called weak localization correction to the system conductance
\cite{AA}. This correction is large provided the electrons moving in the
metal remain coherent. On the other hand, this phase coherence
can persist only for a finite time and is eventually destroyed
due to various processes, such as electron-electron and
electron-phonon interactions, spin-flip scattering, etc. This
characteristic decoherence time $\tau_{\varphi}$ plays a prominent
role in the theory of weak localization \cite{AA,CS}.

In the absence of magnetic impurities and if the temperature of
the system is sufficiently low the decoherence time $\tau_{\varphi}$
is determined by  electron-electron interactions. It was demonstrated in
Ref. \cite{AAK} (see also \cite{CS,SAI}) that for this dephasing mechanism
the decoherence time increases with temperature as
$\tau_{\varphi} \propto T^{2/(d-4)}$, where $d$ is the system dimension.
This theoretical prediction was verified in several experiments \cite{Gio,Pooke}
over a certain temperature interval.

Does the divergence of $\tau_{\varphi}$ in
the zero temperature limit imply that coherence is not
destroyed at $T=0$? Recent experiments \cite{Webb} clearly
suggest a negative answer, indicating that at very low
temperatures the time $\tau_{\varphi}$ saturates at a finite level
showing no tendency for further increase with decreasing $T$.
The authors \cite{Webb} argued that this saturation is not caused by
heating or magnetic impurities but rather is a fundamental consequence
of zero-point fluctuations of electrons. A saturation of $\tau_{\varphi}$
at low $T$ was also observed in earlier works (see e.g. \cite{Gio,Pooke}).

The aim of this paper is to demonstrate that the observed saturation
of $\tau_{\varphi}$ at lowest temperatures \cite{Webb} can be explained
within the existing theory of weak localization
\cite{CS} if one takes into account quantum fluctuations
of the electric field in a disordered conductor.

We essentially follow the analysis elaborated by Chakravarty and Schmid
\cite{CS} and consider the propagation of an electron with 
the kinetic energy $m\dot{\bbox{r}}^2/2$ in a potential
of randomly distributed impurities $U_{imp}(\bbox{r})$. In addition to that
the electron interacts with the fluctuating electric field
$\bbox{ E} (\bbox{r}, t)=-\nabla V (\bbox{r}, t)$
produced by other electrons. These electrons play the role of an effective
environment.

Let us express the propagating electron amplitude in terms of
the Feynman path integral. Within the quasiclassical approximation 
(which is sufficient as long as the elastic mean free path $l$ exceeds 
the Fermi wavelength $p_Fl \gg 1$) the path integral can be replaced by 
the sum over the classical trajectories obeying the equation of motion
\begin{equation}
m\ddot{\bbox{r}}= -\nabla U_{imp}(\bbox{r}) - e \nabla V (\bbox{r}, t)
\label{cl}
\end{equation}
for each realization of random potentials $U_{imp}(\bbox{r})$ and
$V (\bbox{r}, t)$.
Averaging over disordered configurations of impurities \cite{CS}
yields the effective picture of electron diffusion at the scales bigger
than $l$. Fluctuations of the electric field $\nabla V (\bbox{r}, t)$
lead to the phase decoherence. Defining the phase difference between a
classical electron path $\bbox{r}(t')$ and a time reversed path $\bbox{r}(t-t')$
\begin{equation}
\delta \varphi (\bbox{r}, t) =- e\int_0^{t}dt'
[V(\bbox{r}(t'), t')-V(\bbox{r}(t-t'), t')]
\label{delta}
\end{equation}
(which is nonzero provided $V$ fluctuates in space and time)
and averaging with respect to fluctuations of $V$, for
not very small $t$ one gets \cite{CS}
\begin{equation}
\langle (\delta \varphi (\bbox{r},t))^2 \rangle /2 =
t/\tau_{\varphi}(T),
\label{fl}
\end{equation}
where
\begin{equation}
\frac{1}{\tau_{\varphi} (T)}=
\frac{e^2}{a^{3-d}}\int dt\int\frac{d\omega d^dq}{(2\pi)^{d+1}}
\langle|V_{q,\omega}|^2\rangle e^{-Dq^2|t|-i\omega t},
\label{tau}
\end{equation}
$a$ is the film thickness for $d=2$ and $a^2=s$ is the wire cross section
for $d=1$.

The correlation function for voltages in (\ref{tau}) can be
determined with the aid of the fluctuation-dissipation theorem \cite{LL}.
For the sake of definiteness let us consider a quasi-one-dimensional
conductor. Then one finds
\begin{equation}
\langle|V_{q,\omega}|^2\rangle=
\frac{\omega\coth\left(\frac{\omega}{2T}\right)}
{\frac{\omega^2C^2}{\sigma q^2}+\sigma q^2(1+\frac{CD}{\sigma })^2}.
\label{VV}
\end{equation}
Here $\sigma = 2e^2N_0Ds$ is the classical Drude conductance, $D$ is the
diffusion coefficient, and $C$ is the
capacitance of a linear conductor per unit length.
In (\ref{VV}) we neglected retardation and skin effects
which may become important only at very high frequencies. Substituting
(\ref{VV}) into (\ref{tau}) and integrating over $t$ and $q$ after a trivial
algebra we find
\begin{equation}
\frac{1}{\tau_{\varphi}(T)}=\frac{e^2\sqrt{2D}}{\sigma }
\int\limits_{1/\tau_{\varphi}}^{1/\tau_{e}}
\frac{d\omega}{2\pi}\frac{\coth (\omega /2T)}{\sqrt{\omega}}.
\label{tau1}
\end{equation}
In eq. (\ref{tau1}) we made use of the condition $C \ll \sigma /D$ which is
usually well satisfied (perhaps except for extremely thin wires) indicating
the smallness of capacitive effects in our system. Eq. (\ref{tau1}) yields
\begin{equation}
\frac{1}{\tau_{\varphi}}=
\frac{e^2}{\pi\sigma }\sqrt{\frac{2D}{\tau_e}}
\left[2T\sqrt{\tau_e\tau_\varphi} + 1\right].
\label{tau2}
\end{equation}
The first term in the square brackets comes from the low frequency
modes $\omega <T$ whereas the second term is due to high frequency ($\omega >T$)
fluctuations of the electric field in a disordered conductor. At sufficiently high
temperature the first term dominates and the usual expression \cite{AAK}
$\tau_{\varphi} \sim (\sigma /e^2D^{1/2}T)^{2/3}$ is recovered. As $T$ is
lowered the number of the low frequency modes decreases and eventually vanishes
in the limit $T \to 0$. At
$T \lesssim T_q \sim 1/\sqrt{\tau_{\varphi}\tau_e}$ the expression (\ref{tau2})
is dominated by the second term and $\tau_{\varphi}$ saturates at the value
\begin{equation}
\tau_{\varphi} \approx \pi \sigma /e^2v_F
\label{tau4}
\end{equation}
(we disregard the numerical prefactor of order one). The estimate for
the crossover temperature $T_q$ reads
\begin{equation}
T_q \sim ev_F/\sqrt{\sigma l}.
\label{Tq}
\end{equation}
Making use of eq. (\ref{tau4}) it is also easy to find the weak localization
correction $\delta \sigma$ to the Drude conductance in the limit $T=0$.
For $T \lesssim T_q$ we obtain
\begin{equation}
\frac{\delta \sigma }{\sigma}=-\frac{e^2}{\pi \sigma }\sqrt{D\tau_{\varphi}}
\approx - \frac1{p_Fs^{1/2}},
\label{delsig}
\end{equation}
i.e. $\delta \sigma \approx - \sigma /\sqrt{N}$, where $N \sim p_F^2s$ is
the effective number of conducting channels in a 1d mesoscopic system.

For 2d and 3d systems the same analysis yields
\begin{eqnarray}
\frac{1}{\tau_\varphi} & = & \frac{e^2}{4\pi\sigma \tau_e}
[1+2T\tau_e\ln(T\tau_\varphi)],  \; \; \; \; \; \; \;  \quad{\rm 2d}, 
\nonumber \\
\frac{1}{\tau_\varphi} & = & \frac{e^2}
{3\pi^2\sigma\sqrt{2D}\tau_e^{3/2}}[1+6(T\tau_e)^{3/2}],  \quad{\rm 3d},
\label{11}
\end{eqnarray}
where $\sigma =2e^2N_0Da^{3-d}$ is the conductance of a $d$-dimensional
system. The result (\ref{11}) demonstrates that for 2d and 3d systems
saturation of $\tau_{\varphi}$ is expected already at relavitely high
temperatures: the corresponding crossover temperature $T_q$ is of the order
of the inverse elastic time in the 3d case and $T_q \sim v_F/l \ln (p_F^2al)^2$
for a 2d system. The latter value agrees well with the experimental results
\cite{Gio}.
  
The physical origin of the decoherence time saturation
at low temperatures is quite transparent: in the limit $T \to 0$
the dephasing effect is due to
quantum fluctuations of the electric field produced by
electrons in a disordered conductor. This decoherence
effect is by no means surprizing. In fact, it is well known that even at
$T=0$ interaction of a quantum particle with an external quantum bath
leads to the loss of quantum coherence and -- under certain conditions --
to localization of this particle (see e.g. \cite{CLS,SZ}).

Our analysis clearly suggests that at sufficiently low temperatures 
the decoherence time $\tau_{\varphi}$ is {\it not} equal to the
inelastic mean free time $\tau_{i}$, which is known to become infinite at
zero temperature for almost all processes, including electron-electron
interaction. In order to find $\tau_{i}$ it is sufficient to proceed within
the standard quasiclassical approach and to solve the kinetic equation for
the electron distribution function. The collision integral in this equation
contains the product of the occupation numbers for different
energy levels $n_k(1-n_q)$, which vanishes at $T\to  0$ due to the Pauli
principle. As a result $\tau_{i}$ diverges in the zero temperature limit.

In terms of the path integral analysis this procedure amounts to expanding 
the electron effective action on the Keldysh contour in the parameter
$\bbox{r}_{-}(t')=\bbox{r}_1(t')-\bbox{r}_2(t')$ assuming this parameter
to be small ($\bbox{r}_{1(2)}(t')$ is the electron coordinate on the
forward (backward) part of the Keldysh contour). This procedure is formally
very different from one used to calculate the weak localization correction
to conductivity \cite{CS}. In the latter case time reversed pathes 
$\bbox{r}_1(t')$ and $\bbox{r}_2(t-t')$ are assumed to be close to each
other whereas $\bbox{r}_{-}(t')$ can be arbitrarily large. This formal
difference is just an illustration of the well know fact, that  
weak localization is an essentially quantum phenomenon. Therefore, the
standard quasiclassical kinetic analysis of $\tau_i$ in terms of the 
collision integral -- especially at the lowest temperatures -- appears 
to be insuffient for calculation of the decoherence time.

It is also interesting to point out that the expression for the
electron-electron inelastic time $\tau^{ee}_i$ (see e.g. \cite{AA})
is determined by the integral which (apart from an unimportant numerical
prefactor) coincides with the high frequency part ($\omega >T$) of the integral
(\ref{tau},\ref{VV}). In the case of $\tau^{ee}_i$ the integral has
the high frequency cutoff at the electron energy $\epsilon \sim T$, and one
obtains \cite{AA} $1/\tau^{ee} \propto \epsilon^{d/2} \propto T^{d/2}$.
Comparing this expression for $1/\tau^{ee}_i$ with our results for
the inverse decoherence time $1/\tau_{\varphi}$ we arrive at the
conclusion that the former is {\it never} important as compared to the
latter: at high $T>T_q$ the inverse decoherence time is determined 
by the low frequency Nyqist noise $\omega \ll T$, whereas at low 
$T<T_q$ the main contribution to $1/\tau_{\varphi}$ comes from the
high frequency modes of the electric field fluctuations $\omega \gg T$.
In both cases we have $1/\tau_{\varphi} \gg 1/\tau^{ee}_i$.    

We would like to emphasize that our results are obtained within the
standard theoretical treatment of weak localization effects \cite{CS}
combined with the fluctuation-dissipation theorem. One can elaborate a 
more general analysis starting from the microscopic Hamiltonian for
electrons in a disordered metal with Coulomb interaction, introducing
the quantum field $V$ by means of a Hubbard-Stratonovich transformation
(see e.g. \cite{SZ}) and deriving the effective action for one electron
after integration over the remaining electron degrees of freedom
which play the role of the bath. In the quasiclassical limit
$p_Fl \gg 1$ one arrives at the same results as those obtained here.

Note that the decoherence time saturation at low $T$
has been also discussed in a very
recent preprint by Vavilov and Ambegaokar \cite{VA}. These authors describe
the dephasing effect of electromagnetic fluctuations by means of the
effective Caldeira-Leggett bath of oscillators coupled to the electron
coordinate. As compared to our treatment, there are at least two important
differences: (i) the model \cite{VA} does not account for spacial
fluctuations of the electromagnetic field in the sample and (ii) even at
lowest temperatures the authors \cite{VA} treated fluctuations of the bath
as a white noise with temperature $T$ (cf. eq. (11) of Ref. \cite{VA}).
Within this model saturation of the decoherence time
at $T=0$ was obtained only due to the finite sample size: the
corresponding value $\tau_{\varphi}$ \cite{VA} tends to infinity as the
sample length becomes large. In contrast, our results (\ref{tau1}-\ref{delsig})
do not depend on the length of the conductor.

Our result for the quantum decoherence time (\ref{tau4}) also
appears to be different from that presented by Mohanty, Jariwala and Webb
(eq. (2) of Ref. \cite{Webb}). Note, however, that numerical
values for $\tau_{\varphi}$ obtained from our eq. (\ref{tau4})
for the samples Au-1,3,4,6 of \cite{Webb} are in a surprizingly
good agreement with the corresponding estimates derived in Ref. \cite{Webb}.
The latter in turn agree with the experimental data obtained in \cite{Webb}.

Weak localization corrections to the conductance of 1d wires have been
also investigated by Pooke {\it et al.} \cite{Pooke}. At very low
temperatures these authors observed a finite length 
$L_\varphi=\sqrt{D\tau_\varphi}$, which scales as $\sqrt{\sigma}$ 
(with other parameters being fixed) in agreement with our eq. (\ref{tau4}).  

In 2d films the decoherence time saturation at low $T$ was experimentally
found in Ref. \cite{Gio}. The authors attributed this effect to spin-spin
scattering. In our opinion (which seems to be shared by the authors \cite{Gio})
this explanation is not quite satisfactory because
it does not allow to understand the linear dependence of $1/\tau_{\varphi}$
on the sheet resistance of the film detected in \cite{Gio}. 
In contrast, this dependence can be easily explained within the analysis
developed here. The result (\ref{11}) is in a quantitative agreement with 
the experimental findings \cite{Gio}.

In conclusion, we point out that the low temperature saturation of the
electron decoherence time found in recent experiments with mesoscopic
conductors can be explained within the existing theory of weak localization
provided the effect of intrinsic quantum fluctuations of the electric field
is properly accounted for. Our results agree well with the experimental data.

We would like to thank C. Bruder, A. Schmid and G. Sch\"on for
valuable discussions. This work was supported by the Deutsche 
Forschungsgemeinschaft within SFB 195 and by the INTAS-RFBR 
Grant No. 95-1305.

\end{multicols}
\end{document}